\documentclass[12pt]{article}

\usepackage[margin=1in]{geometry}
\usepackage{setspace}
\usepackage{graphicx}
\usepackage{amsthm,amsmath,amssymb}
\usepackage{enumerate}
\usepackage{bbm}
\usepackage{color}
\usepackage{multirow}
\usepackage{url}
\usepackage[labelfont=bf]{caption}
\usepackage{apacite}
\usepackage[authoryear]{natbib}
\newcommand{\diag}{\textrm{diag}}
\newcommand{\eps}{\varepsilon}
\newcommand{\beps}{\boldsymbol{\varepsilon}}

\newcommand{\br}{\boldsymbol{g}}
\newcommand{\bzero}{\boldsymbol{0}}
\newcommand{\bone}{\boldsymbol{1}}

\newcommand{\bmu}{\boldsymbol{\mu}}
\newcommand{\brho}{\boldsymbol{\rho}}

\newcommand{\bphi}{\boldsymbol{\phi}}
\newcommand{\bsigma}{\boldsymbol{\sigma}}

\newcommand{\tr}{\textrm{tr}}

\newcommand{\logdet}{\log \det}

\begin{document}
\pagenumbering{gobble}
\title{Estimating Large Correlation Matrices for International Migration}
\author{Jonathan J. Azose and Adrian E. Raftery\textsuperscript{1} \\
University of Washington}
\date{May 27, 2016}
\footnotetext[1]{Jonathan J. Azose is a Graduate Research Assistant and Adrian E. Raftery is a Professor of Statistics and Sociology, both at the Department of Statistics, Box 354322, University of Washington, Seattle, WA 98195-4322 (Email: jonazose@u.washington.edu/raftery@u.washington.edu). This work was supported by NIH grants R01 HD54511 and R01 HD70936 and by a Science Foundation Ireland ETS Walton visitor award, grant reference 11/W.1/I2079.}
\maketitle 

\newpage

\begin{abstract}
The United Nations is the major organization producing and regularly 
updating probabilistic population projections for all countries. 
International migration is a critical component of such projections, 
and between-country correlations are important
for forecasts of regional aggregates. 
However, there are 200 countries and only 12 data points, each one 
corresponding to a five-year time period. Thus a $200 \times 200$
correlation matrix must be estimated on the basis of 12 data points.
Using Pearson correlations produces many spurious correlations.
We propose a maximum \emph{a posteriori} estimator for the correlation matrix with an interpretable informative prior distribution.
The prior serves to regularize the correlation matrix, shrinking \emph{a priori} untrustworthy elements towards zero. 
Our estimated correlation structure improves projections of net migration for regional aggregates, producing narrower projections of migration for Africa as a whole and wider projections for Europe. 
A simulation study confirms that our estimator outperforms both the Pearson correlation matrix and a simple shrinkage estimator when estimating a sparse correlation matrix. \\

\noindent {\bf Keywords:} Correlation, High-dimensional matrices, 
International Migration, 
World Population Prospects.
\end{abstract}

\newpage
\baselineskip=18pt

\tableofcontents
\pagenumbering{arabic}
\newpage

\listoftables
\listoffigures

\newpage

\section{Introduction}

International migration is a major contributor to population change, but is hard to project, making proper quantification of uncertainty especially important.
Existing global models for migration are well-calibrated marginally,
i.e.~for individual countries \citep{azose2015}, but typically rely on an unrealistic modeling assumption that forecast errors are uncorrelated across countries.
If correlations exist, but are not modeled, the resulting projections may
still be well calibrated for countries individually, but can under- or overestimate variance in projections of migration for regions that span multiple countries.
We present a method for estimating a correlation matrix from a small number
of data points that uses informative priors, shrinking elements of the correlation matrix which we expect \emph{a priori} to be small.
In applying this method to migration, we choose priors based on empirical evidence of non-zero correlations among classes of countries which are ``close'' to one another according to a variety of distance covariates.
Our method improves projections of migration for regional aggregates while mitigating the issue of spurious correlations that arises from trying to estimate a large correlation matrix based on many short time series.

\subsection{Illustrative example}

In this section we focus on six selected countries---Estonia, Latvia, Lithuania, South Africa, Zimbabwe, and Zambia---to highlight the need for regularization of the correlation matrix.

\begin{figure}
\centering
\includegraphics[width=\textwidth]{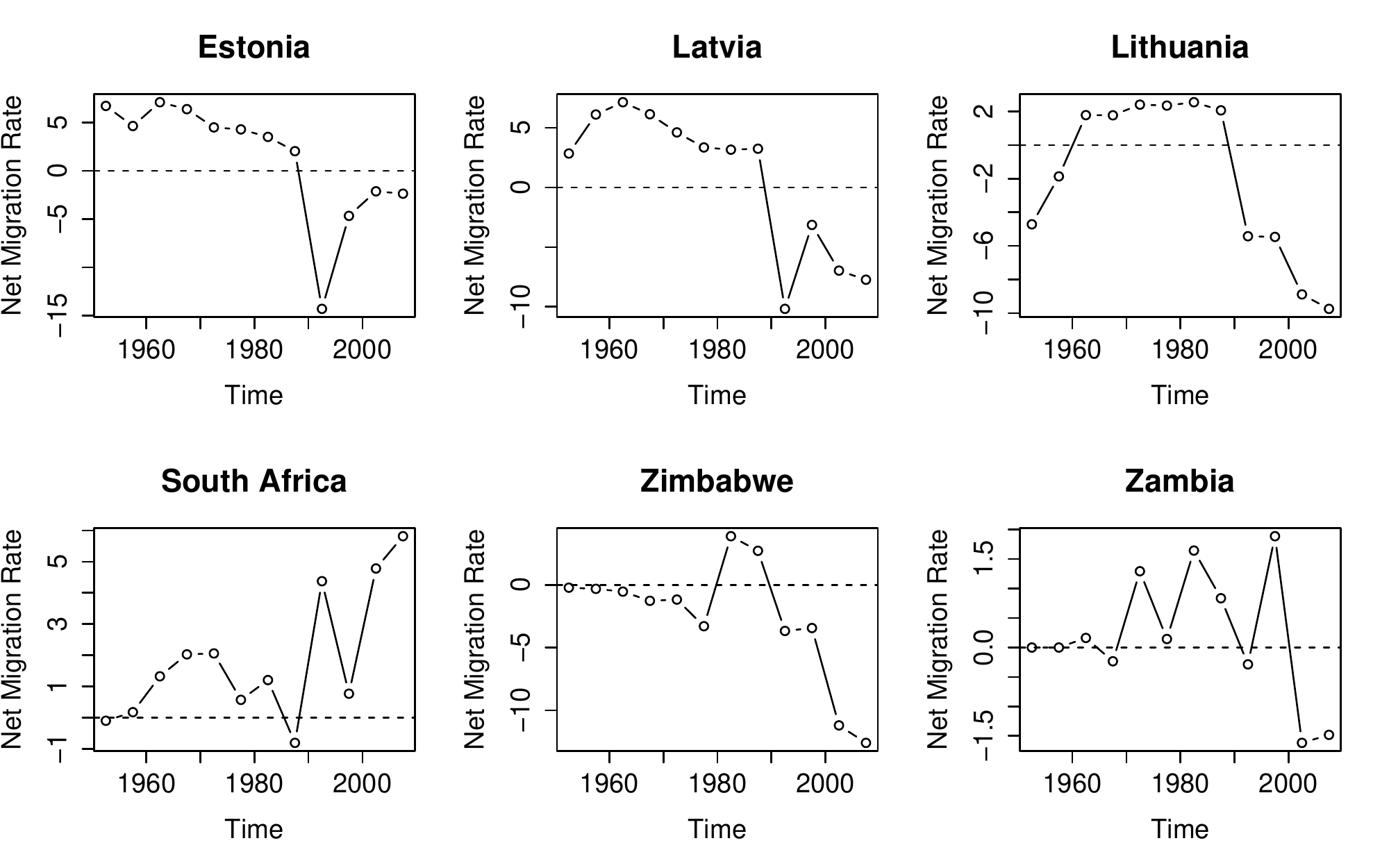}
\caption[Net migration rates for six countries]{Net migration rates (net annual migrants per thousand individuals) for six countries.}\label{fig:sixCountries}
\end{figure}

Migration rates in Estonia, Latvia, and Lithuania over the period from 1950 to 2010 look quite similar (top row of Figure \ref{fig:sixCountries}.)
All three countries share a spike in out-migration during the 1990--1995 time period, which appears as a large negative forecast error in a first-order 
autoregressive (AR(1)) model.
This sudden jump in out-migration among the Baltic states shares a common cause, namely the fall of the Soviet Union, which both induced westward migration and prompted many ethnic Russians to return to Russia \citep{fassmann1994, okolski1998}

Meanwhile, several countries in Southern Africa also experienced big shifts in migration rates during the 1990--1995 time period (bottom row of Figure \ref{fig:sixCountries}.)
From 1990 to 1995, South Africa received substantially more in-migration than it had in previous decades, while Zimbabwe and Zambia both switched from being net receivers of migrants to net senders.
For these three countries, at least some of the change in migration was due to political shifts related to the end of South Africa's apartheid policy.
For example, the number of legal entrants to South Africa who overstayed their visas grew dramatically during the 1990s, with many such entrants coming from other countries of the Southern African Development Community \citep{crush1999}.

\begin{figure}
\centering
\includegraphics[width=\textwidth]{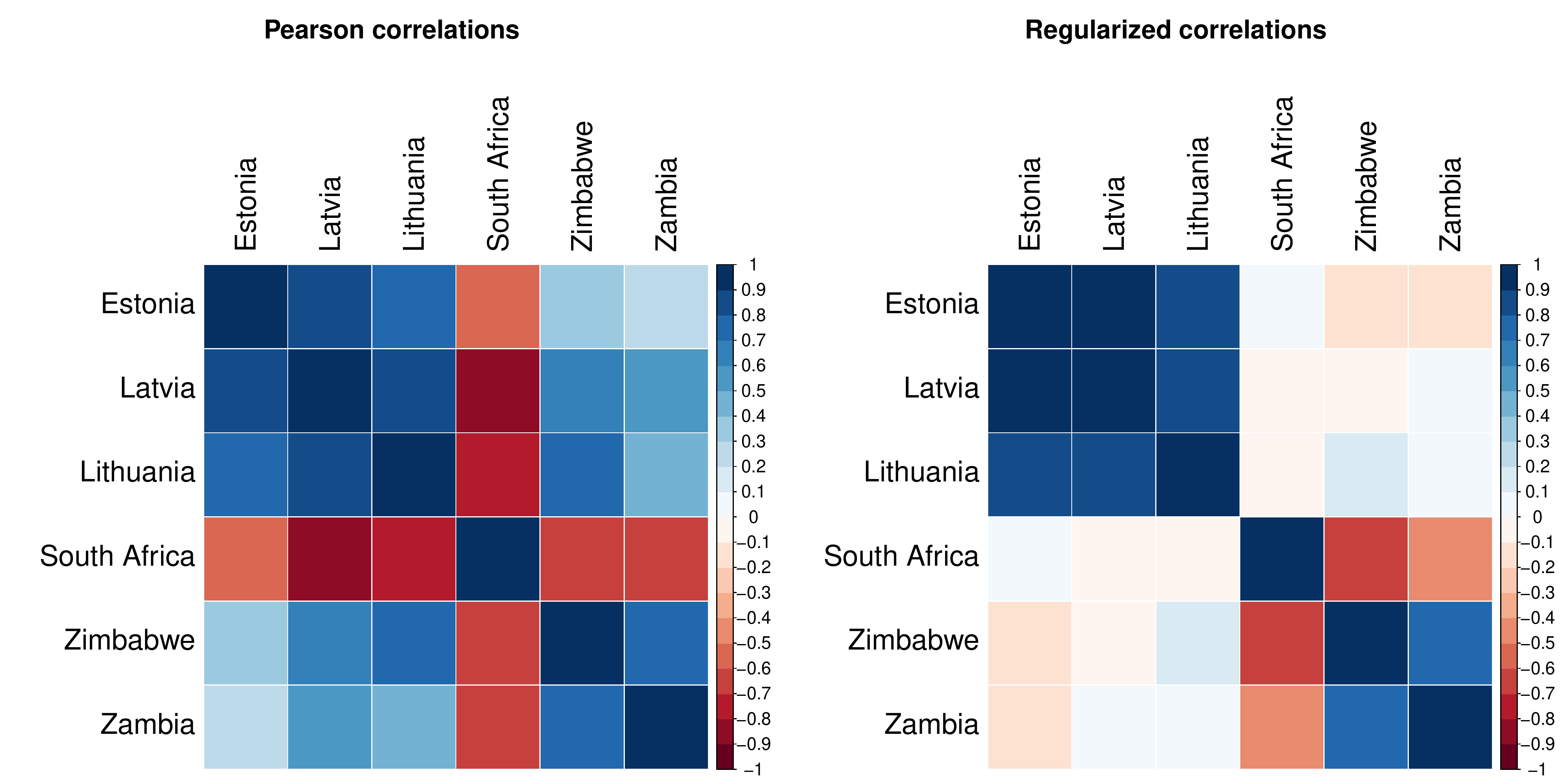}
\caption[Estimated correlations among forecast errors for migration]{Estimated correlations among forecast errors for migration. Left panel shows Pearson correlation estimates. Right panel shows our regularized estimates.}\label{fig:heatmaps}
\end{figure}

Because all six countries experienced pronounced changes in migration rates during the same time period, the usual Pearson estimates of the correlation in forecast errors are relatively large for these six countries (left panel of Figure \ref{fig:heatmaps}.)
Knowledge of world affairs, however, suggests that some of these correlations may be spurious.
There are plausible explanations for the correlations within the three Baltic nations and within the Southern African nations, but the cross-regional correlations are suspect.
In fact, the cross-regional correlations seem to have arisen largely from a coincidental synchrony in the timing of disparate geopolitical events, and do not represent correlations that we would expect to continue to exist in future migration data.
Our method is designed to shrink these seemingly spurious cross-regional correlations, producing the estimated correlation matrix shown in the right panel of Figure \ref{fig:heatmaps}.
Cross-regional correlations decrease substantially in magnitude, while correlations within regions remain largely unchanged.

\subsection{Background}

Country-specific projections of international migration are an important input in policy-making decisions \citep{bijak2007, brownBean2012}.
Projected migration figures are commonly used in long-term planning of social welfare programs \citep{socialSecurity2013, wright2010}.
However, projection of migration is difficult---\cite{bijak2010} describe migration as ``barely predictable''---and global modeling of migration remains somewhat rudimentary.
The United Nations Population Division produces global projections of fertility, mortality, and migration for all countries \citep{wpp2012}.
For most countries, the 2012 revision of the World Population Prospects (WPP) deterministically projects net migration to persist at current levels until 2050 and decline linearly thereafter.

To produce fully probabilistic population projections, one must incorporate probabilistic projections of fertility and mortality with a global probabilistic model of migration. It follows from the demographic balancing equation that the contribution of migration to population change is given by \emph{net} migration (that is, in-migration minus out-migration.) Probabilistic models exist for both net migration \citep{azose2015,azose2016} and in- and out-migration separately \citep{wisniowski2015}. Both of these models are Bayesian hierarchical autoregressive models which treat forecast errors in migration as independent across countries, conditional on model parameters. 
This leads to projections that are well calibrated for individual countries,
but may not be for multi-country aggregates.
Our method aims to relax this independence assumption.

It is worth noting that a strong correlation in migration rates themselves need not translate to a strong correlation in forecast errors.
For example, from 1960 through 2000, Mexico was consistently either the largest or second-largest source of migration flows to the US, with nearly 5 million Mexicans migrating to the US during the 1990's \citep{abel2013}.
While we estimate that net migration rates for the USA and Mexico have a correlation of -0.56 based on quinquennial WPP data from 1950-2010, we estimate a correlation in forecast errors of only -0.07.
That is, most of the relationship between the USA and Mexico is already captured by the autoregressive model parameters, and the ``random'' components of migration rates for the two countries are nearly independent conditional on the AR(1) model.

In this high-dimensional setting with short time series, 
the empirical correlation matrix is a poor estimator, in that it can include many spuriously large estimated correlations.
Our goal is to use regularization to improve an empirical correlation matrix for forecast errors in migration.
There is a large body of literature on regularized estimation of covariance matrices, with applications in genomics, image processing, and finance, among other fields \citep{fan2014}. 
The novelty of our method is that it allows the incorporation of 
available prior information in an easily interpretable way.

Existing covariance estimators based on penalized likelihood maximization are typically maximum {\em a posteriori} (MAP) estimates under some prior belief about covariance, but these formulations are not well suited to specifying beliefs directly about elements of the correlation matrix. Perhaps the most similar method to ours is that of \cite{bien2011}, which allows informative priors on elements of the covariance matrix rather than the correlation matrix. Their method is not directly applicable to our setting, as our goal is to augment existing marginal variances with a suitable correlation structure. Other proposed MAP estimators include the graphical lasso \citep{friedman2008}, which can be used to place an informative prior on the inverse covariance, and the method of \cite{chi2014}, which penalizes covariance estimates that have very large or very small eigenvalues. An extreme example is given by \cite{chaudhuri2007}, who provide a method for covariance estimation in the presence of known zeroes. \cite{zhangZou2014} propose a variant on penalized likelihood maximization that replaces the negative log-likelihood with a simpler loss function.

A related class of covariance estimators relies on shrinkage of an empirical covariance matrix towards a simpler estimator, typically trading some bias for lower mean squared error \citep{ledoit2003,ledoit2004,ledoit2012}. 
A strength of these methods is that so long as the empirical covariance matrix is positive semi-definite and the shrinkage target is positive definite, a linear combination of the two will naturally be positive definite. Applying a shrinkage method to the migration setting would be difficult, as the elements we would like to penalize do not define a positive definite shrinkage target. 

A form of regularization that is straightforward to implement is applying thresholding directly to elements of a covariance or correlation matrix \citep{bickelLevina2008Thresholding, elKaroui2008}; these authors show that a hard-thresholded covariance matrix is consistent in operator norm. Generalized thresholding \citep{antoniadis2001}, developed in the context of wavelet applications, provides a class of related regularized estimators. A key difficulty with such estimators is that care must be taken to ensure that the resulting estimator is positive definite. In some problems, this can be handled by selecting a thresholding constant from an appropriate range \citep{fanLiaoMincheva2013}. Unfortunately, such an approach is not easily adapted to our problem. The structure of the elements we wish to penalize is such that we can tolerate only a small amount of shrinkage of all penalized elements before our estimated correlation matrix loses positive definiteness. 

One fully Bayesian treatment is proposed by \cite{liechty2004}, who include substantive prior information by specifying clusters of correlations which they expect to be similar. This is unfortunately unsuitable to our setting, since geographical and cultural proximity can give rise to either positive or negative correlations. \cite{huang2013} describe a computationally attractive \emph{non-informative} prior on covariances which does not easily extend to the informative priors we would like to include. Other fully Bayesian treatments are given by \cite{barnard2000}, who propose a prior on the correlation matrix  which is either marginally or jointly uniform, and \cite{leonard1992} and \cite{deng2013}, who propose Bayesian estimation of the logarithm of the covariance matrix, which is unfortunately hard to interpret.

In scenarios where there is a natural ordering to the variables, it is often reasonable to make the assumption that large values of $|i-j|$ imply near independence or conditional independence. When this is the case, one can regularize by banding or tapering of the covariance or inverse covariance matrix \citep{bickel2008, fan2007, furrer2007, chen2013, levina2008}. These approaches are not suitable to our problem, as there is no natural ordering of countries.

Good overviews of other methods in covariance estimation are given by \cite{fan2015} and \cite{pourahmadi2011}.

\section{Methods}

We start with an established, well-calibrated autoregressive model on net migration rates for all countries \citep{azose2015}.
This model has the form:
\begin{align}
\br_t - \bmu &= \textrm{diag}(\bphi) (\br_{t-1} - \bmu) + \beps_t,\\
\beps_t &\overset{iid}{\sim} \mathcal{N}_C\left( \bzero, \textrm{diag}(\bsigma) \cdot I_C \cdot \textrm{diag}(\bsigma) \right) , \\
\phi_c &\overset{\textrm{iid}}{\sim} U(0,1) , \\
\mu_c &\overset{\textrm{iid}}{\sim} N(\lambda, \tau^2) , \\
\sigma^2_c &\overset{\textrm{iid}}{\sim} IG(a,b).
%a &\sim U(1,10) , \\
%b|a &\sim U(0,100(a-1)) , \\
%\lambda &\sim U(-100,100) , \\
%\tau &\sim U(0,100).
\end{align}
Notationally, $\br_t$ is a length-$C$ vector of net migration rates for all countries during the time period from $t$ to $t+1$, where $C$ is the number
of countries analyzed.
The quantities $\bmu$, $\bphi$, and $\bsigma$ are vectors of model parameters, and $\bzero$ is a length-$C$ vector of zeroes.
(We have omitted here the specifics of hyperpriors on $a$, $b$, $\lambda$, and $\tau$, which Azose and Raftery selected to reflect the ranges of plausible values.)
Notably, forecast errors in their model are treated as independent, 
conditional on the model's other parameters.
Our method augments this model with an estimated correlation structure.
Although this paper focuses on the migration context, the same technique 
could be applied to probabilistic models of other demographic indicators.

From this point forward, we refer to Azose and Raftery's model as the Bayesian Hierarchical Model with Independent Forecast Errors (BHM+IFE).
In principle, the methodology we describe here provides a means of estimating a correlation matrix to be adjoined to any probabilistic model with conditionally independent forecast errors.

The outline of our procedure for estimating a correlation matrix is as follows:
\begin{enumerate}
\item From the BHM+IFE model, draw a posterior sample of $m$ realizations of model parameters, $\bmu^{(1)}$, $\bphi^{(1)}$, $\bsigma^{(1)}$, \ldots, $\bmu^{(m)}$, $\bphi^{(m)}$, $\bsigma^{(m)}$.
\item Convert the estimated forecast errors from the posterior sample of model parameters to a single empirical correlation matrix, $\tilde R$.
\item Combine the empirical correlation matrix with informative priors on correlations to obtain a maximum \emph{a posteriori} (MAP) correlation estimate, $\hat R$.
\end{enumerate}

This procedure can be viewed as performing a single step of the Monte Carlo EM (MCEM) algorithm \citep{wei1990}. 

The posterior sampling in stage 1 can be performed using any reasonable sampling procedure. In practice, we performed our posterior sampling with a combination of Gibbs sampling and Metropolis-Hastings steps. 

In the following sections, we first discuss the details of obtaining an MAP estimator (Section \ref{sec:MAP}) and then the question of what to use for an empirical correlation matrix (Section \ref{sec:RTilde}).
This is followed by an algorithm for computing the MAP estimator (Section \ref{sec:minimization}), and finally discussion of a criterion for selecting a regularization parameter (Section \ref{sec:lambda}).

\subsection{MAP correlation estimate}\label{sec:MAP}

Our goal is to estimate the correlation structure, $R$, of forecast errors, $\beps_t$. We assume a model of the form
\begin{equation}\label{eqn:normalLikelihood}
\beps_t \overset{iid}{\sim} \mathcal{N}_C\left( \bzero, \Sigma \right),
\end{equation}
where the variance matrix, $\Sigma$, decomposes into standard deviations, $\bsigma$, and a correlation matrix, $R$, as $\Sigma = \textrm{diag}(\bsigma) \cdot R \cdot \textrm{diag}(\bsigma)$.
To determine a MAP estimator for $R$, we express the posterior distribution for $R$ as a product of likelihood and prior. 

\subsubsection{Data Likelihood}

Equation (\ref{eqn:normalLikelihood}) implies a likelihood function for $R$ of the form
\begin{equation}
p(\beps_1, \ldots, \beps_{T-1} | R, \bsigma) \propto_R \det(R)^{-(T-1)/2} \exp \left(
-\frac{1}{2} \sum_{t=1}^{T-1} \beps'_t \diag(\bsigma)^{-1} R^{-1} \diag(\bsigma)^{-1} \beps_t
\right),
\end{equation}
restricted to the space $\Omega$ of valid correlation matrices (i.e.~positive semi-definite matrices with ones on the diagonal.) Matrix trace identities simplify this likelihood to
\begin{equation}
p(\beps_1, \ldots, \beps_{T-1} | R, \bsigma) \propto_R \det(R)^{-(T-1)/2} \exp \left(
-\frac{1}{2} \tr(R^{-1} \tilde R)
\right),
\end{equation}
where
\begin{equation}\label{eqn:RTilde}
\tilde R := \frac{1}{T-1} \sum_{t=1}^{T-1} \diag(\bsigma)^{-1} \beps_t \beps'_t \diag(\bsigma)^{-1}.
\end{equation}
The evidence from the data is encapsulated in $\tilde R$, which is something akin to an empirical correlation matrix.
Note that $\tilde R$ would be a sufficient statistic for $R$ if the $\beps_t$'s and $\bsigma$ were known. 
In fact neither the $\beps_t$'s nor $\bsigma$ are known, and $\tilde R$ must be replaced with a sensible estimate in order to proceed.
Details of the estimation of $\tilde R$ are given in Section \ref{sec:RTilde}.

\subsubsection{Prior}

Our choice of prior distribution on $R$ is motivated by a desire to incorporate informative prior beliefs about which country pairs are likely to be nearly uncorrelated.
As such, we choose a prior of the form
\begin{equation}
\pi(R) \propto_R \prod_{0 \leq i < j \leq C} \exp(-\lambda P_{ij} |R_{ij}|),
\end{equation}
again restricted to $\Omega$. The matrix $P$ with entries $P_{ij}$ is a penalty matrix that encodes the extent to which we believe that countries $i$ and $j$ may be correlated. In our application to migration, we constrain all the entries
in $P$ to be equal to 0 or 1, although in general $P$ may be allowed to have arbitrary non-negative entries. The parameter $\lambda$ is an overall regularization parameter that encodes how strongly we want to penalize correlations.

The key benefit of this prior is its ease of interpretability. Setting $P_{ij}=1$ expresses a belief that $R_{ij}$ should be close to zero, with the strength of that belief controlled by $\lambda$. Setting $P_{ij}=0$ implies that all values of $R_{ij}$ are equally believable, \emph{a priori}. Other penalized likelihood estimators have been proposed, corresponding to MAP estimators under implied priors on precision \citep{friedman2008}, covariance \citep{bien2011}, or eigenvalues of the covariance matrix \citep{chi2014}. 
None of these allow one to specify prior beliefs about correlations directly.

Note that under this specification, the prior distribution of the correlation
$R_{ij}$ is either uniform or truncated Laplace conditional on the rest of the correlation matrix, but marginal distributions will not be uniform or double exponential.
Although it is possible to specify a marginally uniform prior on all elements of the correlation matrix \citep{barnard2000}, we know of no way to specify a distribution that is marginally uniform for some elements and marginally peaked at zero for others.

Because the prior density is a product of Laplace densities on correlations, we will refer to our eventual correlation estimator as the LPoC (Lapalace Prior on Correlations) estimator. Augmenting the BHM+IFE with the LPoC correlation estimate produces the BHM+LPoC model.

\subsubsection{Posterior}

Combining the likelihood and prior, we obtain the log posterior distribution for $R$, equal to
\begin{equation}
\log p(R | \beps_1, \ldots, \beps_{T-1} , \bsigma) = - \frac{T-1}{2}\logdet(R) - \frac{T-1}{2} \tr(R^{-1} \tilde R) - \frac{\lambda}{2}\|P*R\|_1 + c(\beps_1, \ldots, \beps_{T-1}, \bsigma),
\end{equation}
where $*$ denotes elementwise matrix multiplication, and $\| \cdot \|_1$ gives the sum of the absolute value of the elements of a matrix.

Thus, finding the MAP estimator for $R$ is equivalent to solving the minimization problem
\begin{equation}\label{eqn:minProblem}
\textrm{Minimize}_{R \in \Omega}  \left\{
\logdet(R) + \tr(R^{-1} \tilde R) + \frac{1}{T-1} \lambda \cdot \|P*R\|_1.
\right\}
\end{equation}

Algorithmic details of a numerical solution are given in Section \ref{sec:minimization}.

Note that if the penalty parameter, $\lambda$, is zero, then this minimization problem yields the maximum likelihood estimator (MLE)  of R conditional on $\bsigma$. So long as $\tilde R$ is itself positive definite, this MLE is just $\tilde R$, the empirical correlation matrix. Similarly, if $\lambda$ is held fixed as $T$ grows, the penalty term in (\ref{eqn:minProblem}) goes to zero and the LPoC estimator converges to $\tilde R$. Since $\tilde R$ is consistent for $R$, the LPoC estimator is also consistent.

\subsection{Estimating $\tilde R$}\label{sec:RTilde}

Since the forecast errors and model parameters of the BHM+IFE model are unknown, we do not have access to the true value of $\tilde R$.
Instead we use an estimate of $\tilde R$.
For practical reasons, we would prefer to have $\tilde R$ itself be a valid correlation matrix so that (\ref{eqn:minProblem}) will have a known analytic solution in the limiting scenarios where $T$ grows or $\lambda$ goes to zero.
Accordingly, we might choose an estimator $\tilde R^{basic}$ with elements defined by
\begin{equation}
\tilde R_{ij}^{basic} := \frac{\sum_{t=1}^{T-1} \hat\eps_{i,t} \hat\eps_{j,t}}
{\sqrt{\sum_{t=1}^{T-1} \hat\eps_{i,t}^2} \sqrt{\sum_{t=1}^{T-1} \hat\eps_{j,t}^2}},
\end{equation}
where $\hat \beps_t$ is the posterior mean of $\beps_t$ from the BHM+IFE model.
This estimate, $\tilde R^{basic}$, is the MLE for estimating the correlation matrix of a multivariate normal random variable with mean known to be zero and unknown marginal variance terms.
By construction, $\tilde R^{basic}$ is guaranteed to be positive semi-definite and to have ones on the diagonal.

However, in our application, $\tilde R^{basic}$ is low rank, since $T$ is small relative to the dimension of the matrix. For computational reasons, we would prefer to have a strictly positive definite matrix, so we estimate $\tilde R$ by
\begin{equation}
\tilde R^{PD} = 0.99 \cdot \tilde R^{basic} + 0.01 \cdot I_C.
\end{equation}
This change can be viewed as augmenting our estimates of $\beps_t$ with a small amount of additional uncorrelated data.

\subsection{Solving the minimization problem}\label{sec:minimization}

We apply a majorize-minimize algorithm similar to that used by \cite{bien2011} to the minimization problem in (\ref{eqn:minProblem}). 
The function being minimized over is the sum of a convex and a concave component. 
The majorize-minimize algorithm repeatedly iterates through the following steps:
\begin{enumerate}
\item Replace the concave component with its tangent plane to obtain a fully convex function.
\item Find the global minimum of the convex function from Step 1.
\item Update the estimate of the tangent plane.
\end{enumerate}
Notationally, we label our starting point for this algorithm as $R_0$ and subsequent iterations of this majorize-minimize algorithm are denoted with subscripts $R_1, R_2, \ldots$.

In (\ref{eqn:minProblem}), the concave component is $\log \det (R)$, which we replace with the tangent plane $\log \det R_i + \tr(R_i^{-1}(R-R_i))$. 
After simplifying and removing terms which are constant in $R$, the convex minimization problem in the $i$th iteration of the algorithm is
\begin{equation}\label{eqn:innerMinimization}
\textrm{Minimize}_{R \in \Omega} \left\{ \tr (R_i^{-1} R)+ \tr(R^{-1} \tilde R) + \lambda \| P * R\|_1 \right\}.
\end{equation}

Now all of the terms the objective function in (\ref{eqn:innerMinimization}) are convex, and all but $\lambda \|P * R\|_1$ are differentiable, so we can apply the generalized gradient descent algorithm \citep{beck2009}. 
Each generalized gradient descent step takes the form
\begin{equation}\label{eqn:equivalent}
R_{new}=\textrm{argmin}_{\omega \in \Omega} \{(2t)^{-1} \| \omega - (R_{current}- t(R_i^{-1} -R_{current}^{-1} \tilde R R_{current}^{-1}))\|_F^2 + \frac{\lambda}{T-1} \|P * \omega \|_1 \}.
\end{equation}
If the restriction to $\Omega$ were not present, this problem would have a simple analytic solution, given by
\begin{equation}\label{eqn:proposedStep}
R_{new}=\mathcal{S}\left(R_{current}- t(R_i^{-1} -R_{current}^{-1} \tilde R R_{current}^{-1}), \frac{\lambda}{T-1}tP\right),
\end{equation}
where $\mathcal{S}$ is the element-wise soft-thresholding operator defined by
\begin{equation}
\mathcal{S}(X,\alpha)_{ij}=\textrm{sign}(X_{ij}) \cdot (|X_{ij}|-\alpha_{ij}) \cdot \mathbbm{1}(|X_{ij}|>\alpha_{ij}).
\end{equation}
(This move is actually restricted to the off-diagonal elements only, as the diagonal elements of a correlation matrix are constrained to equal 1.)
Thus, if there were no positive definiteness constraint, each update step would consist of a gradient descent step according to the gradient of the differentiable component followed by soft-thresholding the result.

Although we do have to satisfy a positive definiteness constraint, we can start by trying the update step in (\ref{eqn:proposedStep}).
If this update results in a valid correlation matrix, then that matrix is our solution to (\ref{eqn:equivalent}), and we replace $R_{current}$ with $R_{new}$.
However, sometimes the soft-thresholded gradient step results in a matrix that is not positive definite.
In that case, it is possible to appeal to a slower, iterative solution to (\ref{eqn:equivalent}).
One such solution is given by \cite{cui2016}. 
In practice, as long as we are looking for a solution in the interior of $\Omega$, it is good enough to simply reduce step size rather than appealing to the relatively costly iterative algorithm whenever the generalized gradient descent suggestion lies outside of $\Omega$.

Step size selection has a large impact on performance and convergence of this algorithm. Details of step size selection are discussed in Appendix \ref{sec:stepSize}.

\subsection{Selecting the regularization parameter $\lambda$}\label{sec:lambda}

Although the penalty matrix $P$ can be selected on the basis of world knowledge, we are less likely to have genuine prior beliefs about the value of the regularization parameter $\lambda$.
Accordingly, we need some procedure for selecting a value for $\lambda$.
In regularization problems, it is common to select the regularization parameter via cross-validation \citep{bien2011,chi2014,huang2006}.
This approach is too computationally intensive to be feasible for our application.
Among shrinkage estimators, it is common to choose the amount of shrinkage in order to minimize an expected loss function \citep{james1961,ledoit2003}.
However, no suitable analytic result exists that allows us to approximately minimize expected loss in our scenario.

Consequently, we developed a heuristic criterion that selects $\lambda$ in a way that aligns with the goal of our regularization process.
Our method's intent is to shrink the magnitude of penalized elements of the correlation matrix while leaving unpenalized elements more or less unchanged.
In practice, although we succeed at bringing penalized elements towards zero, this shrinkage usually comes at the cost of inflating other elements.
We have observed that this inflation tends to grow more pronounced as $\lambda$ grows.
For very large values of $\lambda$, our estimated correlation matrix may shrink nearly all penalized entries to zero at the expense of inflating a few elements (both penalized and unpenalized) to nearly $\pm 1$. This is not a desirable outcome.

Although it may seem counterintuitive at first,
the observed inflation is not an artifact of a coding error or poor convergence of our algorithm. A simple reproducible example of inflation in a $3 \times 3$ matrix is provided in Appendix \ref{sec:inflation}. In this low-dimensional setting, standard numerical optimization routines agree with the results from our code and both display inflation of unpenalized elements.

Our criterion for selecting $\lambda$ compares the off-diagonal elements of $\tilde R$ and $\hat R(\lambda)$. We choose the value of $\lambda$ which maximizes the difference between average shrinkage and average inflation. Formally, our criterion is defined by
\begin{equation}
k(\tilde R, \lambda) = 
\underset{i,j \textrm{ s.t. } |\hat R(\lambda)_{ij}| < |\tilde R_{ij}|}{\textrm{mean}} \left(|\tilde R_{ij}| - |\hat R(\lambda)_{ij}|\right)
-
\underset{i,j \textrm{ s.t. }|\hat R(\lambda)_{ij}| > |\tilde R_{ij}|}{\textrm{mean}} \left(|\hat R(\lambda)_{ij}| - |\tilde R_{ij}|\right).
\end{equation}
Large positive values of $k$ are desirable, as they correspond to values of $\lambda$ for which we induce a lot of shrinkage and not much inflation.

\section{Results}

In this section,
we first report results from applying our method to global migration data in Section \ref{sec:application}.
Section \ref{sec:simStudy} then provides a simulation study which demonstrates that our method outperforms Pearson correlations and the Ledoit-Wolf shrinkage estimator \citep{ledoit2003} in the scenario where the penalty matrix $P$ is appropriate to the true correlation structure.

\subsection{Application to migration}\label{sec:application}

\subsubsection{Data}

We use data on net migration from the 2012 revision of the World Population Prospects (WPP) \cite{wpp2012}. 
The WPP contains estimates of net migration for all countries in five-year time periods from 1950 until 2010, a total of 12 time periods. We compute the net migration rate $g_{c,t}$ as the net number of migrants in country $c$ over the five year period starting at time $t$, divided by thousands of individuals in country $c$ at time $t$.

Because we want to express prior beliefs as a function of distance covariates, we restrict the set of modeled countries to the 191-country overlap between the WPP 2012 and the set of countries included in CEPII's GeoDist database, a database of bilateral distance covariates defined on pairs of countries \citep{mayer2011}.

\subsubsection{Selection of $P$}

Our estimation technique requires that we choose a penalty matrix, $P$, that reflects our prior beliefs about which country pairs are likely to be correlated. Although it would be possible to elicit expert opinion about each of the roughly 18,000 country pairs, we instead choose a $P$ that can be characterized in terms of just a few covariates. Our matrix $P$ penalizes a pair of countries if \emph{none} of the following conditions is met:
\begin{enumerate}
\item The two countries are contiguous.
\item The two countries' most important cities are located less than 3000 km apart.
\item The two countries are in the same region according to the United Nations 
Population Division's division of the world into 22 regions, based on
both geographical contiguity and cultural affinity \citep{wpp2012}.
\item The two countries are currently in a colonial relationship.
\end{enumerate}
This definition of $P$ is in line with migration theory, which suggests that migrant flows are more likely when monetary and social costs of movement are low \citep{harris&1970, lee1966, sjaastad1962, stark&1985}, as will be the case with countries which are geographically proximate or share administrative ties.
This definition penalizes 85\% of country pairs, leaving 15\% unpenalized. The average country is considered to be ``close'' to 29 other countries, and ``distant'' from the remaining 161. 

In selecting these conditions, we examined nine candidate distance covariates. 
The first eight such covariates come from CEPII's GeoDist database \citep{mayer2011}, while the ninth is derived from the United Nations division 
into 22 regions.
The left column of Table \ref{tab:ksTest} gives the complete list of covariates considered. As an empirical basis for determining which criteria to include in defining our penalty matrix, we examined the elements of the sample correlation matrix for all pairs of countries meeting each criterion. Using a Kolmogorov-Smirnov test, we tested whether the distribution of these sample correlations was different from the distribution of elements of the sample correlation matrix under a null hypothesis of uncorrelated errors. The right column of Table \ref{tab:ksTest} shows the $p$-values from these Kolmogorov-Smirnov tests. Our definition of the penalty matrix $P$ includes all covariates with a $p$-value less than 0.05.

\begin{table}
\caption[Results of Kolmogorov-Smirnov test for correlations]{Results of Kolmogorov-Smirnov test that empirical correlations are significantly different from the distribution of elements of a sample correlation matrix when the true error structure is uncorrelated. $p$-values lower than 0.05 are bolded.}\label{tab:ksTest}
\begin{center}
\begin{tabular}{rc}
\hline
Covariate & $p$-value\\
\hline
Contiguous & \textbf{0.019} \\
Common language (official) & 0.23 \\
Common language (spoken by 9\% of pop.) & 0.58\\
Geodesic distance less than 3000 km & \textbf{0.0003}\\
Colonial relationship after 1945 & 0.57\\
Common colonizer after 1945 & 0.11\\
Current colonial relationship & \textbf{0.035}\\
Ever had a colonial link & 0.36\\
Same UN Region & \textbf{0.036}\\
\hline
\end{tabular}
\end{center}
\end{table}

\subsubsection{Selection of the regularization parameter, $\lambda$}

\begin{figure}
\centering
\includegraphics[width=0.6\textwidth]{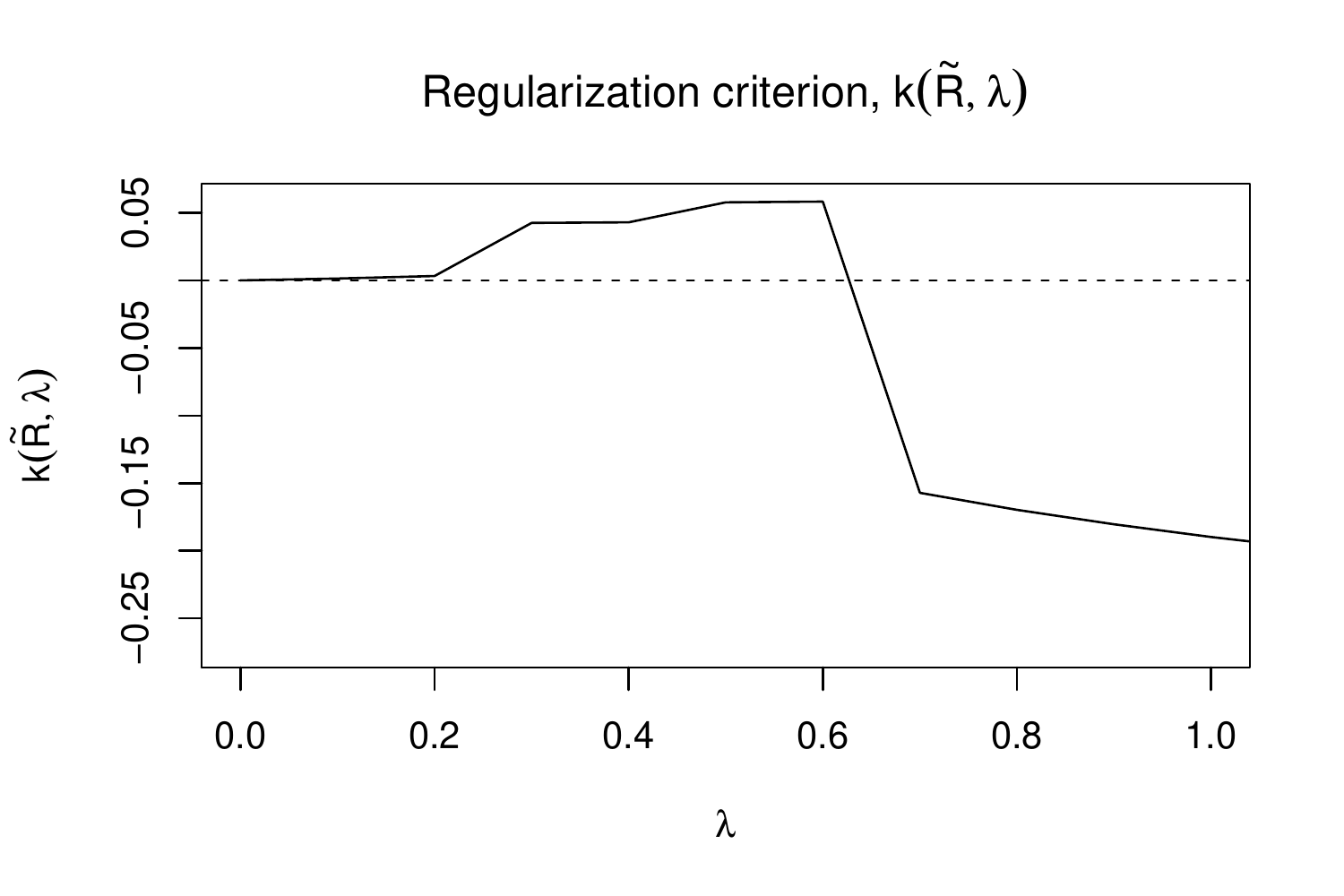}
\caption[Regularization criterion as a function of $\lambda$]{Regularization criterion, $k(\tilde R, \lambda)$ as a function of $\lambda$. The regularization criterion is the difference between the average shrinkage among shrunk elements of $\hat R(\lambda)$ and average inflation among inflated elements.}\label{fig:shrinkageCriterion}
\end{figure}

We computed values of $\hat R(\lambda)$ for all values of $\lambda$ from 0 to 3 in increments of 0.1.
Figure \ref{fig:shrinkageCriterion} shows the value of $k(\tilde R, \lambda)$ over a range of $\lambda$ values. 
We found that $k(\tilde R, \lambda)$ peaked at $\lambda=0.6$, where we find average shrinkage of 0.13 compared with average inflation of 0.07. 
Increasing $\lambda$ from 0.6 to 0.7 induces additional shrinkage, but at the cost of greatly inflating some correlations. 
Accordingly, we choose $\hat R(0.6)$ as our estimate of $R$.

\begin{figure}
\centering
\includegraphics[width=0.6\textwidth]{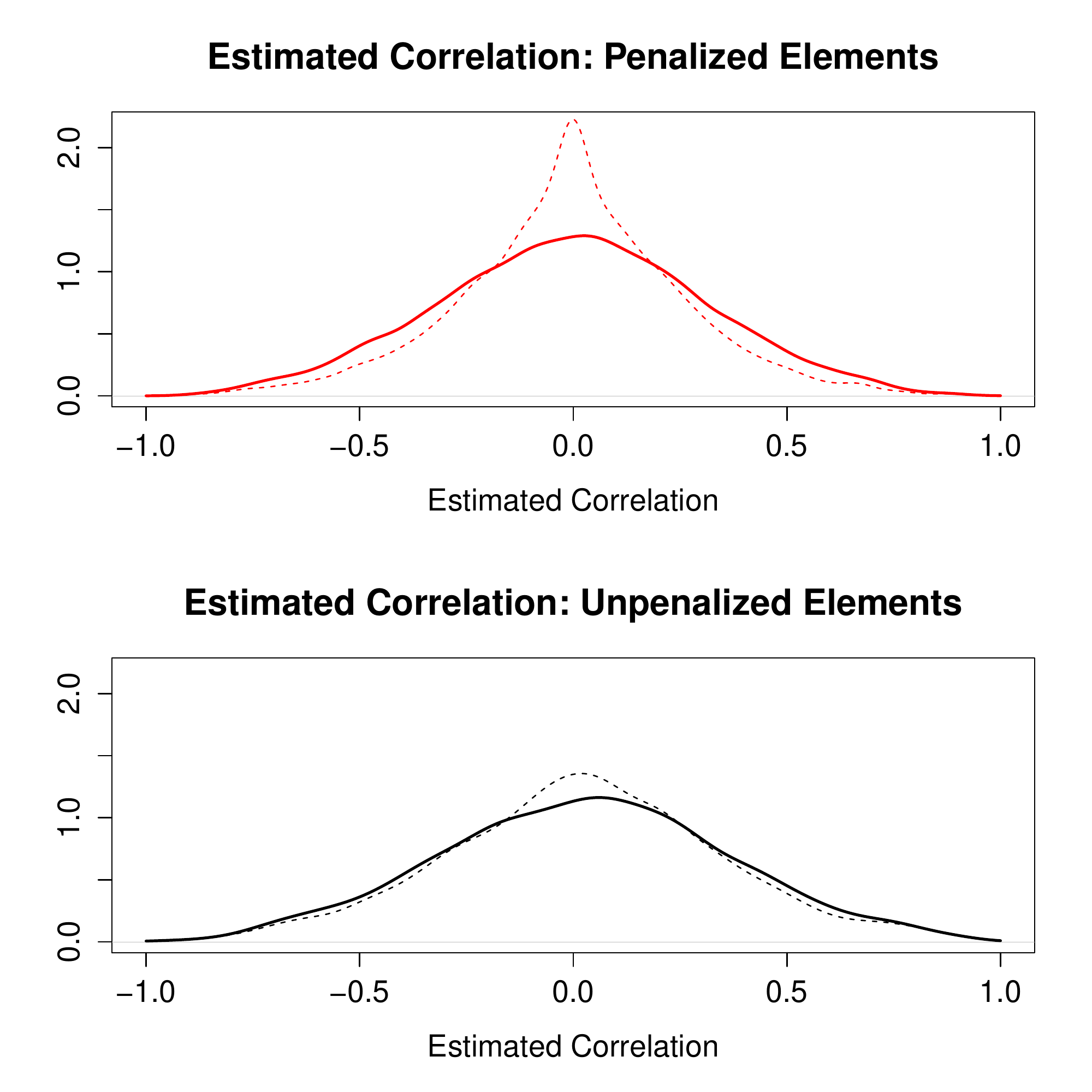}
\caption[Elements of the correlation matrix before and after regularization]{Comparison of elements of the correlation matrix before regularization (solid curves) and after (dashed curves). Top panel shows penalized elements; bottom panel shows unpenalized elements.}\label{fig:splitPlot}
\end{figure}

Figure \ref{fig:splitPlot} shows the impact of regularization on the correlation matrix. 
Among penalized elements (top panel), we see significant shrinkage towards zero, although many penalized elements remain large in magnitude, even after regularization.
The bottom panel shows the unpenalized elements of the correlation matrix before regularization (solid curve) and after (dashed curve). 
On average we induce some shrinkage in the unpenalized elements, but the distribution is largely unchanged. 

\subsubsection{Projection and evaluation}

\begin{figure}
\centering
\includegraphics[width=\textwidth]{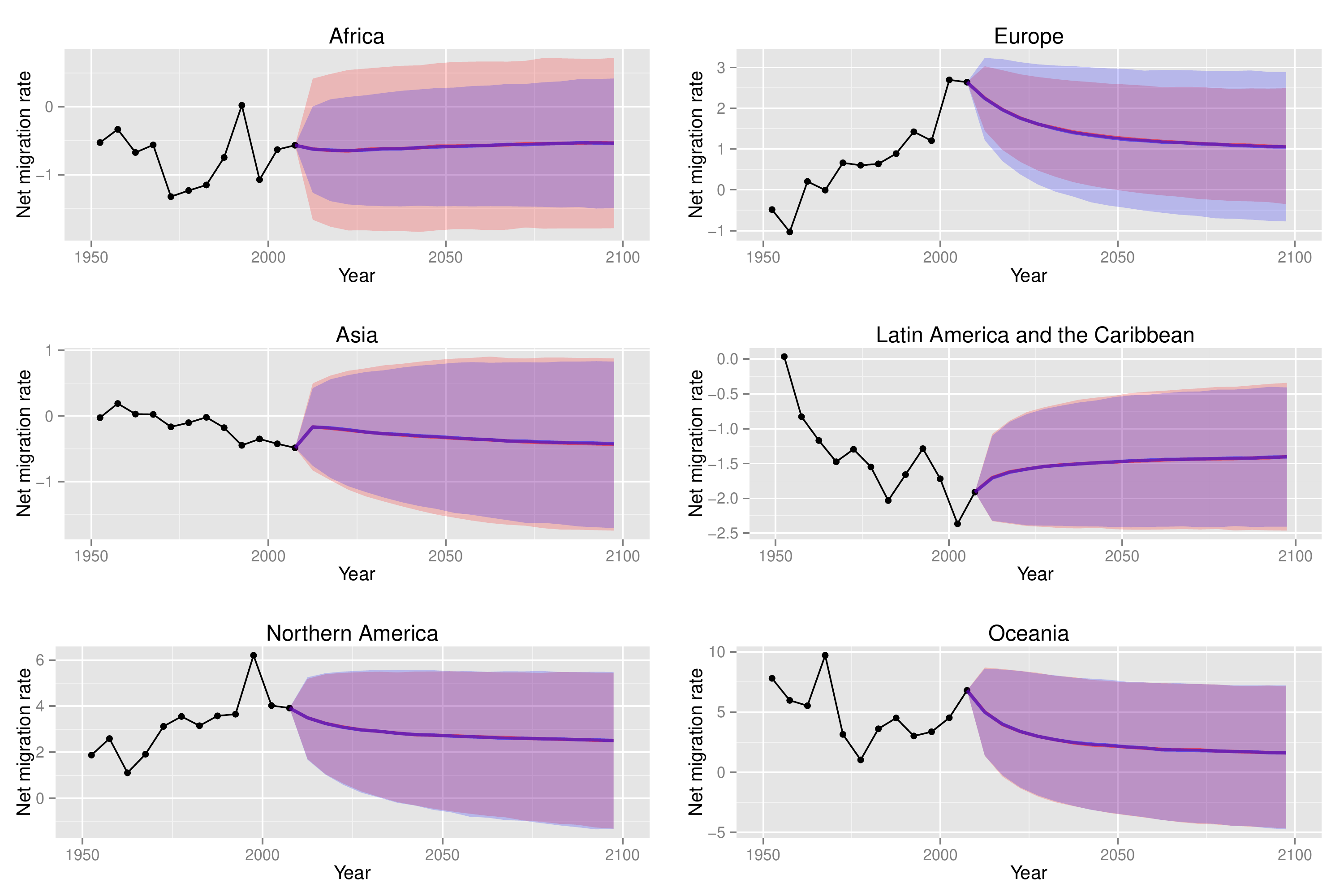}
\caption[Probabilistic projections of net migration for continents]{Medians and 80\% prediction intervals for net migration among continents. Projections from the Bayesian hierarchical model with independent forecast errors (BHM+IFE) are given in red. Projections using our estimated correlation matrix (BHM+LPoC) are in blue. Overlap is in purple.}\label{fig:continents}
\end{figure}

We augment the BHM+IFE model with the LPoC estimate $\hat R(0.6)$ to produce probabilistic projections of migration for any collection of countries.
Figure \ref{fig:continents} contains medians and 80\% prediction intervals of projected migration for all continents.
In Africa, negative correlations narrow our projections. 
In Europe, positive correlations cause forecasts to widen.
For the other continents, we see little change in projected migration.

For evaluation, we compare true migration rates for regional aggregates in 1995--2010 with projections of the same regional aggregates based only on migration data from 1950--1995.
This procedure entails re-estimation of the BHM+IFE model using only the 1950--1995 data, followed by construction of an empirical correlation matrix, selection of $\lambda$, and extraction of $\hat R(\lambda)$.
We compare the performance of the BHM+IFE model on regional aggregates to a model using the same sampled values of $\bmu$, $\bphi$, and $\bsigma$, but augmented with $\hat R(\lambda)$.

As an evaluation metric, we use the negatively oriented continuous ranked probability score (CRPS) \citep{hersbach2000,gneiting2007}.
The CRPS compares the cumulative distribution function, $F$, of a probabilistic forecast to an observation, $x$, and is defined by
\begin{equation}
\mathrm{CRPS}(F,x) = \int_{- \infty}^{\infty} (F(y)- \mathbbm{1}\{y \geq x\})^2\ dy.
\end{equation}
In our application the two probabilistic forecasts under consideration have the same mean as one other, by design.
One approximate way of looking at CRPS in this setting is that when $g_{c,t}$ is close to the mean of the forecast, we reward $F$ for having low variance; when $g_{c,t}$ is far from the mean, we reward $F$ for having high variance.

\begin{figure}
\centering
\includegraphics[width=\textwidth]{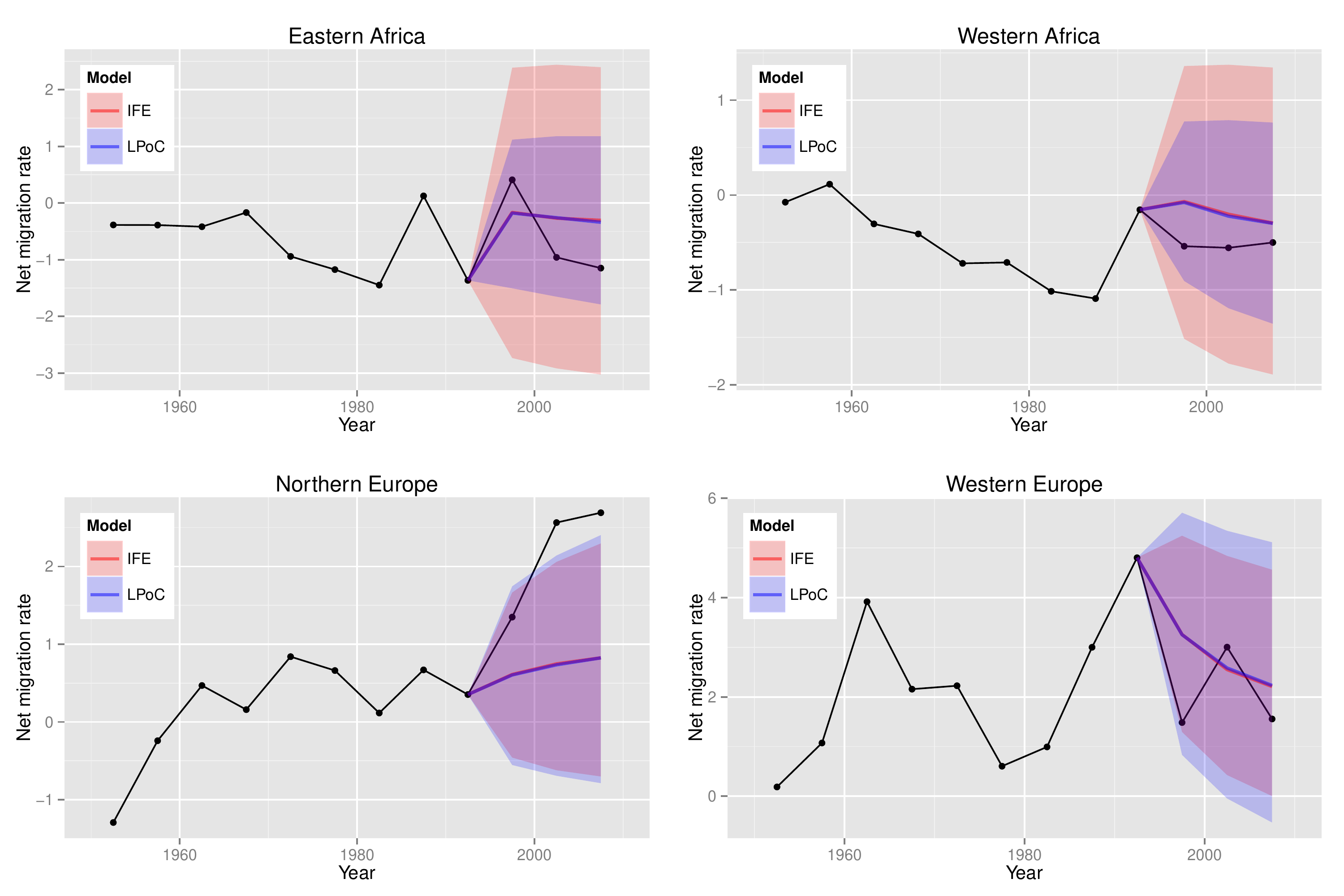}
\caption[Probabilistic projections of net migration for regional aggregates]{Medians and 80\% prediction intervals for projections of net migration rates for regional aggregates. Projections from the Bayesian hierarchical model with independent forecast errors (BHM+IFE) are given in red. Projections using our estimated correlation matrix (BHM+LPoC) are in blue. Overlap is in purple.}\label{fig:fourRegions}
\end{figure}

Table \ref{tab:CRPS} gives CRPS for projections of aggregate migration for the six continents. Our model improves the quality of projections in Africa and Europe, while projections for the other four continents are more or less unchanged.
Figure \ref{fig:fourRegions} illustrates the change in projections of net migration in 1995--2010 for four subregions of Africa and Europe.
Projections from the BHM+IFE model are in red; projections from BHM+LPoC are in blue. 
Our method narrows prediction intervals in Eastern and Western Africa, bringing the width of the 80\% prediction intervals more into line with the range of observed variability.
In both regions, true migration rates for the projected period stayed within our narrower intervals.
In contrast, our method widens projections in Northern and Western Europe, where the 80\% intervals from the BHM+IFE model either miss or nearly miss capturing some of the observed data points.

\begin{table}[ht]
\begin{center}
\caption[Continuous ranked probability score for continental migration projections]{Continuous ranked probability score for all continents evaluated on projections of 1995-2010, where lower is better. 
Left column: Projections based on the Bayesian hierarchical model with independent correlation structure (BHM+IFE). Right column: Projections based on the Bayesian hierarchical model with our regularized correlation estimate (BHM+LPoC). Bolded entry in each row indicates the lower value.}\label{tab:CRPS}
\begin{tabular}{rrrr}
  \hline
 & IFE & LPoC \\ 
  \hline
Africa & 1.66 & \textbf{1.49} \\ 
  Asia & \textbf{0.73} & 0.74 \\ 
  Europe & 3.92 & \textbf{3.76} \\ 
  Latin America and the Caribbean & \textbf{1.62} & \textbf{1.62} \\ 
  Northern America & 5.02 & \textbf{4.99} \\ 
  Oceania & 8.53 & \textbf{8.49} \\ 
   \hline
\end{tabular}
\end{center}
\end{table}

\clearpage

\subsection{Simulation study}\label{sec:simStudy}
In this section we show by simulation that our regularization procedure improves correlation estimates in a low-dimensional setting. 
To match the application of interest, we simulate 12 observed time points from an AR(1) process with correlated errors. 
For computational tractability, we decrease the number of simulated countries from 191 in the real data to 9 in the simulation. For each of 100 simulations, we perform the following procedure:
\begin{enumerate}
\item Generate a set of simulated migration rates $\br_1, \ldots, \br_{12}$ from an AR(1) process with errors correlated as described below.
\item Produce point estimates of $\beps_1, \ldots, \beps_{11}$ via MCMC sampling of $\bmu$, $\bphi$, and $\bsigma$.
\item Convert $\beps_t$'s to a matrix $\tilde R$ using the procedure in Section \ref{sec:RTilde}.
\item Solve the minimization problem (\ref{eqn:minProblem}) to obtain a regularized estimate for the correlation matrix.
\end{enumerate}

Since the procedure for selecting $\lambda$ is computationally intensive, we perform this procedure only once and use the same value of $\lambda$ for all subsequent simulations.

\subsubsection{Simulation details}

We simulate a collection of nine countries with true migration rates governed by the AR(1) process
\begin{equation}
\br_t - \bmu = \textrm{diag}(\bphi) (\br_{t-1} - \bmu) + \beps_t.
\end{equation}
For simplicity we take $\bmu = \bzero$, $\bphi = \frac{1}{2}\bone$, and
\begin{equation}
\beps_t \overset{iid}{\sim} \mathcal{N}_9(\bzero, \Sigma).
\end{equation}
We fix $\Sigma$ to be block diagonal. Compound symmetric correlation structure within each $3 \times 3$ block is given by
\begin{equation}
\Sigma_{3 \times 3}=\left(
\begin{array}{ccc}
1 & 0.5 & 0.5 \\
0.5 & 1 & 0.5 \\
0.5 & 0.5 & 1 
\end{array}
,
\right)
\end{equation}
and the full covariance matrix by
\begin{equation}
\Sigma = \left(
\begin{array}{ccc}
\Sigma_{3 \times 3} & \bzero & \bzero \\
\bzero & \Sigma_{3 \times 3} & \bzero \\
\bzero & \bzero & \Sigma_{3 \times 3} \\
\end{array}
\right).
\end{equation}
We then simulate observations $\br_1, \ldots, \br_{12}$ and attempt to make inference on the correlation structure of $\Sigma$.

\begin{figure}
\centering
\includegraphics[width=0.6\textwidth]{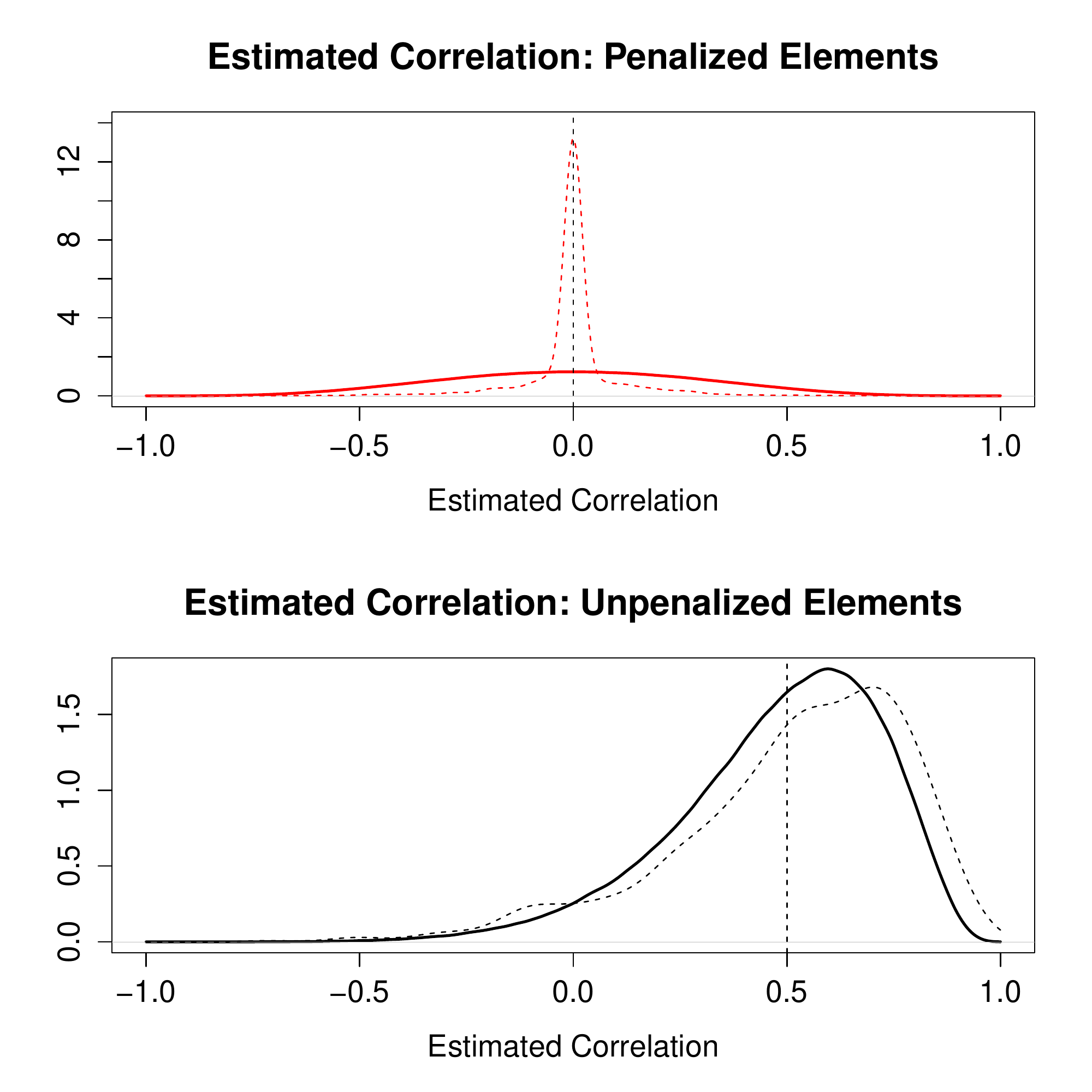}
\caption[Simulation study: Elements of the correlation matrix before and after regularization]{Simulation study results: Comparison of elements of the correlation matrix before regularization (solid curves) and after (dashed curves). Top panel shows penalized elements; bottom panel shows unpenalized elements. True correlations are indicated with dashed vertical lines.}\label{fig:simStudyResults}
\end{figure}

Because we are basing inference on a small number of time points, Pearson estimates of correlation are highly variable.
Solid curves in Figure \ref{fig:simStudyResults} show the distributions of the off-diagonal elements of the unregularized Pearson correlation matrix in the ideal scenario where the values of $\beps_t$ can be perfectly estimated.
The top panel shows the distribution of the elements for which the true correlation is zero. 
The bottom panel shows elements for which the true correlation is 0.5. 
In both cases, high variability makes inference difficult. 
Our method is designed to decrease variability among estimated correlations for those country pairs where prior knowledge suggests that correlation should be close to zero.

To illustrate a best case scenario, we choose a penalty matrix $P$ which is well suited to the true correlation structure. The simplest such $P$ is the one which penalizes the off-diagonal elements of the correlation matrix if and only if the true correlation is zero. That $P$ is given by
\begin{equation}
P=\left(
\begin{array}{ccc}
\bzero_{3 \times 3} & \bone_{3 \times 3} & \bone_{3 \times 3}\\
\bone_{3 \times 3} & \bzero_{3 \times 3} & \bone_{3 \times 3} \\
\bone_{3 \times 3} & \bone_{3 \times 3} & \bzero_{3 \times 3} \\
\end{array}
\right).
\end{equation}

\subsubsection{Initial run to select $\lambda$}

\begin{figure}
\centering
\includegraphics[width=0.6\textwidth]{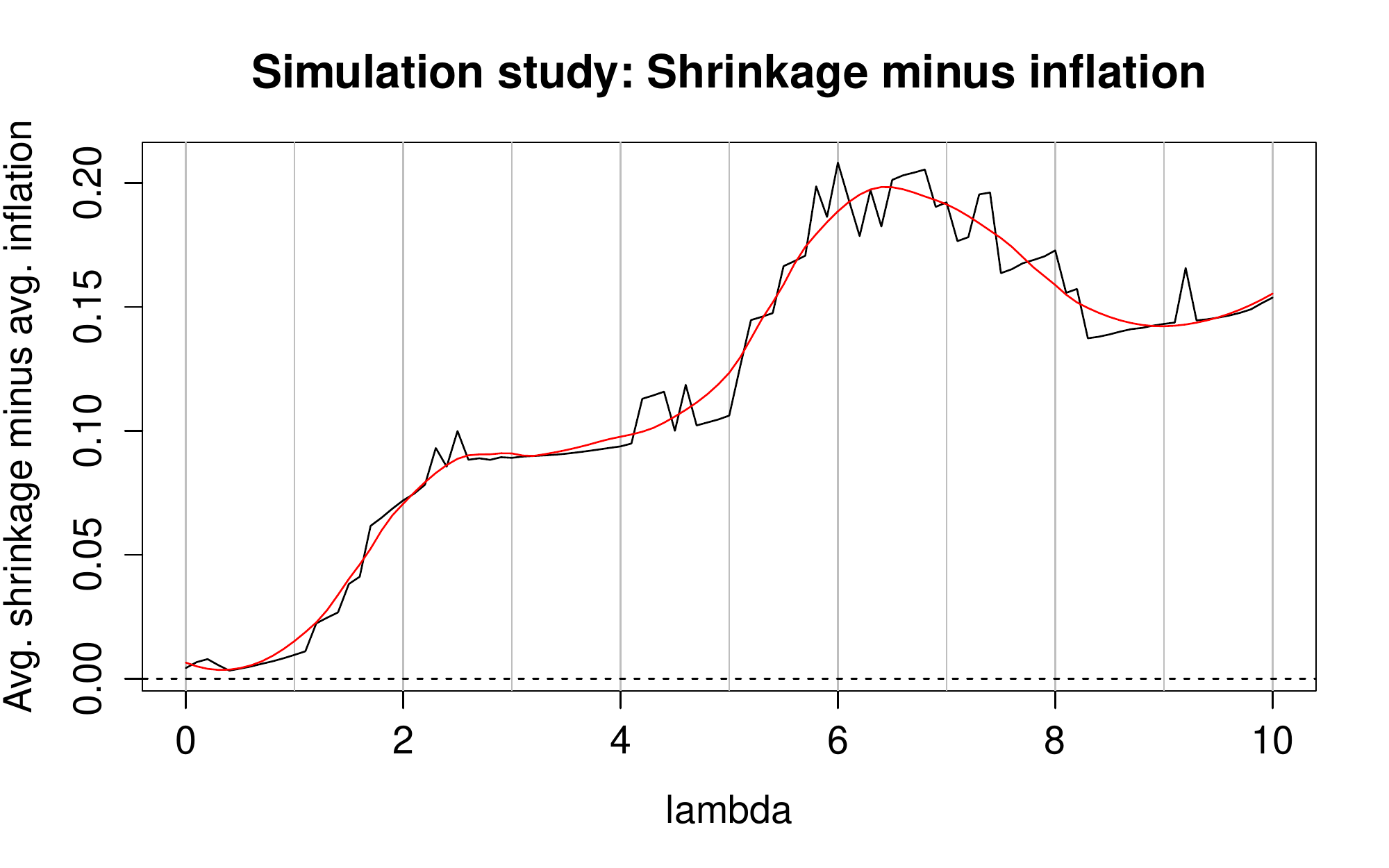}
\caption[Simulation study: Regularization criterion]{Average shrinkage minus average inflation of elements of $\hat R(\lambda)$ as $\lambda$ varies from 0 to 10. Exact curve in black, Lowess-smoothed curve in red.}\label{fig:shrinkage_simulated}
\end{figure}

Our procedure to select $\lambda$ is computationally expensive, as it requires us to compute $\hat R(\lambda)$ repeatedly as $\lambda$ varies. We therefore perform this procedure only once and use the same $\lambda$ for estimation of $R$ in all subsequent simulated data sets. Figure \ref{fig:shrinkage_simulated} plots our $\lambda$-selection criterion based on a single simulated data set over the range $\lambda=0, 0.1, 0.2, \ldots, 10$. The exact curve, shown in black, exhibits some jumpiness in this low-dimensional setting, a problem which naturally becomes less severe in the high-dimensional setting of interest. Because of this jumpiness, we base our selection of $\lambda$ on a Lowess-smoothed curve, selecting the maximizing value of $\lambda=6.4$.

\subsubsection{Evaluation of repeated estimation of $R$}

We produced 100 estimates of $\hat R(\lambda=6.4)$ from 100 different sets of simulated migration rates, all using the same block diagonal correlation structure. Dashed lines in Figure \ref{fig:simStudyResults} show the distribution of off-diagonal elements of $\hat R$, split into those elements where the true correlation is 0 and elements where the true correlation is 0.5 (top and bottom panel, respectively). 

Our method is successful in shrinking penalized elements towards zero. Among elements where the true correlation is zero, we correctly estimate an exact zero in 62\% of cases in this simulation.
Among unpenalized elements, our method produces estimates with slightly more variability (the standard deviation is 0.256 for Pearson correlations versus 0.272 for our estimates).
Both methods produce estimates for unpenalized elements that are within two standard errors of the true mean value of 0.5. The mean estimated correlation is 0.489 for Pearson correlations (standard error 0.009) versus 0.514 for our estimates (standard error of 0.009).
On the whole, the LPoC estimator greatly improves estimates of penalized elements at the expense of slightly increasing variability in unpenalized elements.

Table \ref{tab:simulationEvaluation} compares mean absolute error and mean squared error from our method with two competing estimators. We compare our results against both Pearson correlation matrices and correlation matrices that have been regularized using the Ledoit-Wolf method, which shrinks Pearson estimates towards a spherical correlation structure \citep{ledoit2003}. 
In the top panel, we estimate $\beps_1, \ldots, \beps_{11}$ with a Bayesian hierarchical model, as is done in our real application to migration. 
In the bottom panel, we assume instead a scenario where we have direct access to $\beps_1, \ldots, \beps_{11}$, as would be suitable in other applications where the interest is in estimating correlations of directly observed quantities.
In both cases, our method provides an overall reduction in mean squared error by at least two thirds when compared against the Pearson sample correlation matrix. 
A large reduction in error from shrinking penalized elements is offset by a mild increase in error among unpenalized elements. 
We also outperform the Ledoit-Wolf estimator in terms of overall error.

\begin{table}
\caption[Evaluation of correlation matrix estimates from simulation study]{Evaluation of correlation matrix estimates from simulation study. ``LPoC'' refers to our estimator, which uses Laplace priors on correlations. MAE is mean absolute error. MSE is mean squared error. Averages over ``all elements'' exclude diagonal elements, which are fixed at zero by all methods.
The lowest (best) values are shown in bold.}\label{tab:simulationEvaluation}
\begin{center}
\begin{tabular}{cccc}
\multicolumn{4}{c}{Values of $\beps_t$ estimated with MCMC} \\
& Estimator & MAE & MSE\\
\hline
\hline
\multirow{3}{*}{All elements}
& Pearson & 0.253 & 0.098\\
& Ledoit-Wolf & 0.193 & 0.055\\
& LPoC & \textbf{0.090} & \textbf{0.028}\\
\hline
\multirow{3}{*}{True correlation = 0} 
& Pearson & 0.270 & 0.109\\
& Ledoit-Wolf & 0.190 & 0.053\\
& LPoC & \textbf{0.049} & \textbf{0.012}\\
\hline
\multirow{3}{*}{True correlation = 0.5} 
& Pearson & 0.201 & 0.066\\
& Ledoit-Wolf & \textbf{0.200} & \textbf{0.060}\\
& LPoC & 0.214 & 0.074\\
\hline \hline
\noalign{\vskip 2mm}    
\multicolumn{4}{c}{True values of $\beps_t$ used} \\
& Estimator & MAE & MSE\\
\hline
\hline
\multirow{3}{*}{All elements} 
& Pearson & 0.227 & 0.079\\
& Ledoit-Wolf & 0.182 & 0.047\\
& LPoC & \textbf{0.078} & \textbf{0.022}\\
\hline
\multirow{3}{*}{True correlation = 0} 
& Pearson & 0.244 & 0.089\\
& Ledoit-Wolf & 0.162 & 0.039 \\
& LPoC & \textbf{0.041} & \textbf{0.010}\\
\hline
\multirow{3}{*}{True correlation = 0.5} 
& Pearson & \textbf{0.176} & \textbf{0.051}\\
& Ledoit-Wolf & 0.243 & 0.073\\
& LPoC & 0.190 & 0.058\\
\hline \hline
\end{tabular}
\end{center}
\end{table}

\section{Discussion}

Our method augments probabilistic projections of migration 
that are well-calibrated for individual countries,
with a correlation structure that reflects prior knowledge of between-country correlations.
By combining a high-dimensional empirical correlation matrix with an informative prior that shrinks spurious correlations, we produce an estimated correlation matrix that is in line with migration theory and improves projections of regional aggregates.
When compared with a simple model that assumes uncorrelated forecast errors, our method narrows projections of net migration for Africa and widens projections for Europe.
Out-of-sample evaluation confirms that these changes produce better probabilistic forecasts as measured by continuous ranked probability score.
Mechanically, the novelty of our method is our prior on correlations, which benefits from being interpretable and simple in form, and converts MAP estimation to an $\ell_1$-penalized regularization problem which is computationally tractable.

Our analysis focuses on modeling net migration, but an attractive alternative would be to model a full matrix of bilateral migration flows.
Such a model would naturally imply correlations in migration---if out-migrants from country $i$ tend to go to country $j$, then net migration in countries $i$ and $j$ will be negatively correlated.
However, modeling the global bilateral flow matrix is currently not feasible.
Flows are hard to estimate, even in countries with good data \citep{deBeer2010, raymer2011}.
\cite{abel2013} produces global estimates of migration flows based on migrant stock data, but for only a small number of time periods at which migrant stock data exist.
His method involves minimizing the total number of migrants subject to
the available data on migrant stocks.
This induces many structural zeroes in his estimates, making modeling difficult.
Because of the lack of good data on migration flows, we choose instead to work with net migration rates.

Although our method produces a MAP estimator in the presence of informative priors, we are not able to leverage any of the usual Bayesian machinery to produce a sample from the posterior distribution.
While it would in theory be possible to use MCMC methods to produce a posterior sample by updating one element of the correlation matrix at a time, an updating procedure would need to iterate through some 18,000 elements of the correlation matrix, checking for positive definiteness after each proposed step. 
Such an algorithm is therefore likely to move around the parameter space too slowly to be of any use.
In some settings a Laplace approximation centered at the posterior mode can provide a good approximation of marginal posterior distributions \citep{tierney1986}.
However, the double-exponential priors in our setting render this procedure impracticable.
Within each orthant of the parameter space, a quadratic approximation to the log likelihood is reasonable, but because of the $\ell_1$ penalty term, a different quadratic approximation is required for each of the roughly $2^{18,000}$ orthants, which is not feasible.

Given our interest in combining data with prior beliefs, an inverse Wishart prior on covariance is tempting because it allows easy sampling from the full posterior.
However, the inverse Wishart distribution is restrictive in form \citep{barnard2000} and does not provide a straightforward way to describe prior beliefs about correlations.

Another tempting alternative is that of \cite{liuWangZhao2014}, who give a simple thresholding method for producing a penalized correlation matrix that is guaranteed to be positive definite. Their estimator solves
\begin{equation}
\underset{\omega \succ \delta \cdot I}{\textrm{argmin}} 
\frac{1}{2} \|\tilde R - \omega\|_F^2 + \lambda \|W * \omega\|_{1,\textrm{off}},
\end{equation}
to produce an estimator among the set of valid correlation matrices with minimum eigenvalue no smaller than $\delta$. Although the weight matrix, $W$, is in principle arbitrary, they use $W$ to induce greater shrinkage where empirical correlations are weakest, not as a means of conveying prior information.
We would be hesitant to replace $W$ with our penalty matrix $P$, as that off-license use of their method would not incorporate prior information in a principled way.

Our method can be generalized to shrink estimated correlations towards non-zero values by replacing the penalty term $\lambda\|P*R\|_1$ with $\lambda\|P*(R-S)\|_1$ for some target matrix $S$.
This may be desirable in cases where heavily structured estimates of correlations are available, as is the case for modeling of fertility 
\citep{fosdick2014}.

Note that we have used the 2012 revision of the WPP here \citep{wpp2012}. 
The more recent 2015 revision \citep{wpp2015} contains one additional data point.
It would be of interest to redo the analysis with the newer data, but we expect the results would be similar.

\bibliography{correlationBibliography}

\begin{thebibliography}{}

\bibitem[\protect\astroncite{Abel}{2013}]{abel2013}
Abel, G. (2013).
\newblock Estimating global migration flow tables using place of birth data.
\newblock {\em Demographic Research}, 28:505--546.

\bibitem[\protect\astroncite{Antoniadis and Fan}{2001}]{antoniadis2001}
Antoniadis, A. and Fan, J. (2001).
\newblock Regularization of wavelet approximations.
\newblock {\em Journal of the American Statistical Association},
  96(455):939--955.

\bibitem[\protect\astroncite{Azose and Raftery}{2015}]{azose2015}
Azose, J.~J. and Raftery, A.~E. (2015).
\newblock Bayesian probabilistic projection of international migration.
\newblock {\em Demography}, 52(5):1627--1650.

\bibitem[\protect\astroncite{Azose et~al.}{2016}]{azose2016}
Azose, J.~J., {\v{S}}ev{\v{c}}{\'\i}kov{\'a}, H., and Raftery, A.~E. (2016).
\newblock Probabilistic population projections with migration uncertainty.
\newblock {\em Proceedings of the National Academy of Sciences}.
\newblock Published ahead of print May 23, 2016, doi: 10.1073/pnas.1606119113.

\bibitem[\protect\astroncite{Barnard et~al.}{2000}]{barnard2000}
Barnard, J., McCulloch, R., and Meng, X.-L. (2000).
\newblock Modeling covariance matrices in terms of standard deviations and
  correlations, with application to shrinkage.
\newblock {\em Statistica Sinica}, 10(4):1281--1312.

\bibitem[\protect\astroncite{Beck and Teboulle}{2009}]{beck2009}
Beck, A. and Teboulle, M. (2009).
\newblock A fast iterative shrinkage-thresholding algorithm for linear inverse
  problems.
\newblock {\em SIAM Journal of Imaging Sciences}, 2(1):183--202.

\bibitem[\protect\astroncite{Bickel and
  Levina}{2008a}]{bickelLevina2008Thresholding}
Bickel, P.~J. and Levina, E. (2008a).
\newblock Covariance regularization by thresholding.
\newblock {\em The Annals of Statistics}, 36(6):2577--2604.

\bibitem[\protect\astroncite{Bickel and Levina}{2008b}]{bickel2008}
Bickel, P.~J. and Levina, E. (2008b).
\newblock Regularized estimation of large covariance matrices.
\newblock {\em The Annals of Statistics}, 36(1):199--227.

\bibitem[\protect\astroncite{Bien and Tibshirani}{2011}]{bien2011}
Bien, J. and Tibshirani, R.~J. (2011).
\newblock Sparse estimation of a covariance matrix.
\newblock {\em Biometrika}, 98(4):807.

\bibitem[\protect\astroncite{Bijak et~al.}{2007}]{bijak2007}
Bijak, J., Kupiszewska, D., Kupiszewski, M., Saczuk, K., and Kicinger, A.
  (2007).
\newblock Population and labour force projections for 27 {European} countries,
  2002--2052: {Impact} of international migration on population ageing.
\newblock {\em European Journal of Population}, 23(1):1--31.

\bibitem[\protect\astroncite{Bijak and Wi{\'s}niowski}{2010}]{bijak2010}
Bijak, J. and Wi{\'s}niowski, A. (2010).
\newblock Bayesian forecasting of immigration to selected {European} countries
  by using expert knowledge.
\newblock {\em Journal of the Royal Statistical Society: Series A (Statistics
  in Society)}, 173(4):775--796.

\bibitem[\protect\astroncite{Brown and Bean}{2012}]{brownBean2012}
Brown, S.~K. and Bean, F.~D. (2012).
\newblock Population growth.
\newblock In Gans, J., Replogle, E.~M., and Tichenor, D.~J., editors, {\em
  Debates on {U.S.} {Immigration}}. SAGE, Thousand Oaks, California.

\bibitem[\protect\astroncite{Chaudhuri et~al.}{2007}]{chaudhuri2007}
Chaudhuri, S., Drton, M., and Richardson, T.~S. (2007).
\newblock Estimation of a covariance matrix with zeros.
\newblock {\em Biometrika}, 94(1):199--216.

\bibitem[\protect\astroncite{Chen et~al.}{2013}]{chen2013}
Chen, X., Xu, M., Wu, W.~B., et~al. (2013).
\newblock Covariance and precision matrix estimation for high-dimensional time
  series.
\newblock {\em The Annals of Statistics}, 41(6):2994--3021.

\bibitem[\protect\astroncite{Chi and Lange}{2014}]{chi2014}
Chi, E.~C. and Lange, K. (2014).
\newblock Stable estimation of a covariance matrix guided by nuclear norm
  penalties.
\newblock {\em Computational Statistics \& Data Analysis}, 80:117--128.

\bibitem[\protect\astroncite{Crush}{1999}]{crush1999}
Crush, J. (1999).
\newblock Fortress {South Africa} and the deconstruction of {Apartheid's}
  migration regime.
\newblock {\em Geoforum}, 30(1):1--11.

\bibitem[\protect\astroncite{Cui et~al.}{2016}]{cui2016}
Cui, Y., Leng, C., and Sun, D. (2016).
\newblock Sparse estimation of high-dimensional correlation matrices.
\newblock {\em Computational Statistics \& Data Analysis}, 93:390--403.

\bibitem[\protect\astroncite{De~Beer et~al.}{2010}]{deBeer2010}
De~Beer, J., Raymer, J., Van~der Erf, R., and Van~Wissen, L. (2010).
\newblock Overcoming the problems of inconsistent international migration data:
  A new method applied to flows in {Europe}.
\newblock {\em European Journal of Population}, 26(4):459--481.

\bibitem[\protect\astroncite{Deng and Tsui}{2013}]{deng2013}
Deng, X. and Tsui, K.-W. (2013).
\newblock Penalized covariance matrix estimation using a matrix-logarithm
  transformation.
\newblock {\em Journal of Computational and Graphical Statistics},
  22(2):494--512.

\bibitem[\protect\astroncite{{El Karoui}}{2008}]{elKaroui2008}
{El Karoui}, N. (2008).
\newblock Operator norm consistent estimation of large-dimensional sparse
  covariance matrices.
\newblock {\em The Annals of Statistics}, 36(6):2717--2756.

\bibitem[\protect\astroncite{Fan et~al.}{2014}]{fan2014}
Fan, J., Han, F., and Liu, H. (2014).
\newblock Challenges of big data analysis.
\newblock {\em National Science Review}, 1(2):293--314.

\bibitem[\protect\astroncite{Fan et~al.}{2007}]{fan2007}
Fan, J., Huang, T., and Li, R. (2007).
\newblock Analysis of longitudinal data with semiparametric estimation of
  covariance function.
\newblock {\em Journal of the American Statistical Association},
  102(478):632--641.

\bibitem[\protect\astroncite{Fan et~al.}{2015}]{fan2015}
Fan, J., Liao, Y., and Liu, H. (2015).
\newblock An overview on the estimation of large covariance and precision
  matrices.
\newblock {\em arXiv preprint arXiv:1504.02995}.

\bibitem[\protect\astroncite{Fan et~al.}{2013}]{fanLiaoMincheva2013}
Fan, J., Liao, Y., and Mincheva, M. (2013).
\newblock Large covariance estimation by thresholding principal orthogonal
  complements.
\newblock {\em Journal of the Royal Statistical Society: Series B (Statistical
  Methodology)}, 75(4):603--680.

\bibitem[\protect\astroncite{Fassmann and Munz}{1994}]{fassmann1994}
Fassmann, H. and Munz, R. (1994).
\newblock European {East-West} migration, 1945-1992.
\newblock {\em International Migration Review}, 28(3):520--538.

\bibitem[\protect\astroncite{Fosdick and Raftery}{2014}]{fosdick2014}
Fosdick, B.~K. and Raftery, A.~E. (2014).
\newblock Regional probabilistic fertility forecasting by modeling
  between-country correlations.
\newblock {\em Demographic Research}, 30(35):1011.

\bibitem[\protect\astroncite{Friedman et~al.}{2008}]{friedman2008}
Friedman, J., Hastie, T., and Tibshirani, R. (2008).
\newblock Sparse inverse covariance estimation with the graphical lasso.
\newblock {\em Biostatistics}, 9(3):432--441.

\bibitem[\protect\astroncite{Furrer and Bengtsson}{2007}]{furrer2007}
Furrer, R. and Bengtsson, T. (2007).
\newblock Estimation of high-dimensional prior and posterior covariance
  matrices in {Kalman} filter variants.
\newblock {\em Journal of Multivariate Analysis}, 98(2):227--255.

\bibitem[\protect\astroncite{Gneiting and Raftery}{2007}]{gneiting2007}
Gneiting, T. and Raftery, A.~E. (2007).
\newblock Strictly proper scoring rules, prediction, and estimation.
\newblock {\em Journal of the American Statistical Association},
  102(477):359--378.

\bibitem[\protect\astroncite{Harris and Todaro}{1970}]{harris&1970}
Harris, J.~R. and Todaro, M.~P. (1970).
\newblock Migration, unemployment and development: {A} two-sector analysis.
\newblock {\em American Economic Review}, 60(1):126--142.

\bibitem[\protect\astroncite{Hersbach}{2000}]{hersbach2000}
Hersbach, H. (2000).
\newblock Decomposition of the continuous ranked probability score for ensemble
  prediction systems.
\newblock {\em Weather and Forecasting}, 15(5):559--570.

\bibitem[\protect\astroncite{Huang et~al.}{2013}]{huang2013}
Huang, A., Wand, M.~P., et~al. (2013).
\newblock Simple marginally noninformative prior distributions for covariance
  matrices.
\newblock {\em Bayesian Analysis}, 8(2):439--452.

\bibitem[\protect\astroncite{Huang et~al.}{2006}]{huang2006}
Huang, J.~Z., Liu, N., Pourahmadi, M., and Liu, L. (2006).
\newblock Covariance matrix selection and estimation via penalised normal
  likelihood.
\newblock {\em Biometrika}, 93(1):85--98.

\bibitem[\protect\astroncite{James and Stein}{1961}]{james1961}
James, W. and Stein, C. (1961).
\newblock Estimation with quadratic loss.
\newblock In {\em {Proceedings} of the {Fourth} {Berkeley} {Symposium} on
  {Mathematical} {Statistics} and {Probability}}, volume~1, pages 361--379.

\bibitem[\protect\astroncite{Ledoit and Wolf}{2003}]{ledoit2003}
Ledoit, O. and Wolf, M. (2003).
\newblock Improved estimation of the covariance matrix of stock returns with an
  application to portfolio selection.
\newblock {\em Journal of Empirical Finance}, 10(5):603--621.

\bibitem[\protect\astroncite{Ledoit and Wolf}{2004}]{ledoit2004}
Ledoit, O. and Wolf, M. (2004).
\newblock A well-conditioned estimator for large-dimensional covariance
  matrices.
\newblock {\em Journal of Multivariate Analysis}, 88(2):365--411.

\bibitem[\protect\astroncite{Ledoit and Wolf}{2012}]{ledoit2012}
Ledoit, O. and Wolf, M. (2012).
\newblock Nonlinear shrinkage estimation of large-dimensional covariance
  matrices.
\newblock {\em The Annals of Statistics}, 40(2):1024--1060.

\bibitem[\protect\astroncite{Lee}{1966}]{lee1966}
Lee, E.~S. (1966).
\newblock A theory of migration.
\newblock {\em Demography}, 3(1):47--57.

\bibitem[\protect\astroncite{Leonard and Hsu}{1992}]{leonard1992}
Leonard, T. and Hsu, J.~S. (1992).
\newblock Bayesian inference for a covariance matrix.
\newblock {\em The Annals of Statistics}, 20(4):1669--1696.

\bibitem[\protect\astroncite{Levina et~al.}{2008}]{levina2008}
Levina, E., Rothman, A., Zhu, J., et~al. (2008).
\newblock Sparse estimation of large covariance matrices via a nested lasso
  penalty.
\newblock {\em The Annals of Applied Statistics}, 2(1):245--263.

\bibitem[\protect\astroncite{Liechty et~al.}{2004}]{liechty2004}
Liechty, J.~C., Liechty, M.~W., and M{\"u}ller, P. (2004).
\newblock Bayesian correlation estimation.
\newblock {\em Biometrika}, 91(1):1--14.

\bibitem[\protect\astroncite{Liu et~al.}{2014}]{liuWangZhao2014}
Liu, H., Wang, L., and Zhao, T. (2014).
\newblock Sparse covariance matrix estimation with eigenvalue constraints.
\newblock {\em Journal of Computational and Graphical Statistics},
  23(2):439--459.

\bibitem[\protect\astroncite{Mayer and Zignago}{2011}]{mayer2011}
Mayer, T. and Zignago, S. (2011).
\newblock Notes on {CEPII's} distances measures: The {GeoDist} database.

\bibitem[\protect\astroncite{Nocedal and Wright}{2006}]{nocedal2006}
Nocedal, J. and Wright, S. (2006).
\newblock {\em Numerical {Optimization}}.
\newblock Springer Science \& Business Media.

\bibitem[\protect\astroncite{Ok{\'o}lski}{1998}]{okolski1998}
Ok{\'o}lski, M. (1998).
\newblock Regional dimension of international migration in {Central} and
  {Eastern Europe}.
\newblock {\em Genus}, 54(1):11--36.

\bibitem[\protect\astroncite{Pourahmadi}{2011}]{pourahmadi2011}
Pourahmadi, M. (2011).
\newblock Covariance estimation: The glm and regularization perspectives.
\newblock {\em Statistical Science}, 26(3):369--387.

\bibitem[\protect\astroncite{Raymer et~al.}{2011}]{raymer2011}
Raymer, J., de~Beer, J., and van~der Erf, R. (2011).
\newblock Putting the pieces of the puzzle together: Age and sex-specific
  estimates of migration amongst countries in the {EU/EFTA}, 2002--2007.
\newblock {\em European Journal of Population}, 27(2):185--215.

\bibitem[\protect\astroncite{Sjaastad}{1962}]{sjaastad1962}
Sjaastad, L.~A. (1962).
\newblock The costs and returns of human migration.
\newblock {\em Journal of Political Economy}, 70(5):80--93.

\bibitem[\protect\astroncite{Stark and Bloom}{1985}]{stark&1985}
Stark, O. and Bloom, D.~E. (1985).
\newblock The new economics of labor migration.
\newblock {\em American Economic Review}, 75(2):173--178.

\bibitem[\protect\astroncite{Tierney and Kadane}{1986}]{tierney1986}
Tierney, L. and Kadane, J.~B. (1986).
\newblock Accurate approximations for posterior moments and marginal densities.
\newblock {\em Journal of the American Statistical Association},
  81(393):82--86.

\bibitem[\protect\astroncite{{United Nations}}{2012}]{wpp2012}
{United Nations} (2012).
\newblock {\em World {Population} {Prospects}: The 2012 {Revision}}.
\newblock United Nations, New York.

\bibitem[\protect\astroncite{{United Nations}}{2015}]{wpp2015}
{United Nations} (2015).
\newblock {\em World {Population} {Prospects}: The 2015 {Revision}}.
\newblock United Nations, New York.

\bibitem[\protect\astroncite{{U.S. Social Security
  Administration}}{2013}]{socialSecurity2013}
{U.S. Social Security Administration} (2013).
\newblock {\em {The 2013 Annual Report of the Board of Trustees of the Federal
  Old-age and Survivors Insurance and Federal Disability Insurance Trust
  Funds}}.
\newblock {Board of Trustees, Federal Old-Age and Survivors Insurance and
  Federal Disability Insurance Trust Funds}.

\bibitem[\protect\astroncite{Wei and Tanner}{1990}]{wei1990}
Wei, G.~C. and Tanner, M.~A. (1990).
\newblock A {Monte Carlo} implementation of the {EM} algorithm and the poor
  man's data augmentation algorithms.
\newblock {\em Journal of the American Statistical Association},
  85(411):699--704.

\bibitem[\protect\astroncite{Wi{\'s}niowski et~al.}{2015}]{wisniowski2015}
Wi{\'s}niowski, A., Smith, P.~W., Bijak, J., Raymer, J., and Forster, J.~J.
  (2015).
\newblock Bayesian population forecasting: Extending the {Lee-Carter} method.
\newblock {\em Demography}, 52(3):1035--1059.

\bibitem[\protect\astroncite{Wright}{2010}]{wright2010}
Wright, E. (2010).
\newblock 2008-based national population projections for the {United Kingdom}
  and constituent countries.
\newblock {\em Population Trends}, 139(1):91--114.

\bibitem[\protect\astroncite{Zhang and Zou}{2012}]{zhangZou2014}
Zhang, T. and Zou, H. (2012).
\newblock Sparse precision matrix estimation via lasso penalized {D-trace}
  loss.
\newblock {\em Biometrika}, 99(1):1--18.

\end{thebibliography}

\clearpage

\appendix{}

\section{Determining step size}\label{sec:stepSize}

Step size selection is necessary in high dimensions for the general gradient descent algorithm to converge quickly enough to be useful. Complex methods for step size selection are available, but we obtained reasonable results with the backtracking line search algorithm, which starts with a large step size and decreases step size whenever a proposed step results in too little improvement in the objective function.

Say we have an objective function $f(x)$ which we are trying to minimize. The core of the backtracking line search algorithm is as follows \citep{nocedal2006}.

\begin{enumerate}
\item Fix a backtracking coefficient $\beta \in (0,1)$, a starting step size $\alpha_0$, and a starting location $x_0$.
\item Propose a step of length $\alpha_k$ in direction $p_k$. (The backtracking line search algorithm is a generic algorithm that will work regardless of how the direction $p_k$ is determined.)
\item If the improvement in the objective function is enough to meet the Armijo condition given in  (\ref{eqn:Armijo}) below, then take the proposed step. That is, take $x_{k+1} = x_k + \alpha_k p_k$. Keep the step size constant (i.e., $\alpha_{k+1}=\alpha_k$).
\item Otherwise, if there is any improvement in the objective function, take the proposed step, but also decrease the step size for the next iteration (specifically, set $\alpha_{k+1}=\beta \alpha_k$).
\item Otherwise, there must have been no improvement in the objective function. Don't take a step, but do decrease step size. ($x_{k+1}=x_k$ and $\alpha_{k+1}=\beta \alpha_k$.)
\item Repeat steps 2-5 until convergence.
\end{enumerate}

The Armijo condition, which is used to determine whether to decrease step size, is as follows. The Armijo condition is met if the following inequality is satisfied:
\begin{equation}\label{eqn:Armijo}
f(x_k+\alpha_k p_k) \leq f(x_k) + c_1 \alpha_i \nabla f_k^T p_k.
\end{equation}
($c_1$ is a constant chosen from $(0,1)$ that controls how strictly the change in $f$ must match the gradient at $x_k$.)

In our application, there's a missing component---we can't actually compute the gradient of our objective function. The relevant objective function is given by
\begin{equation}
f(R)=\tr (R_i^{-1} R)+ \tr(R^{-1} \tilde R) + \lambda \| P * R\|_1.
\end{equation}
The first two terms in the sum are differentiable, but the third is not. 

We rewrite the Armijo condition as
\begin{equation}
f(x_k+\alpha_k p_k) \leq f(x_k) + c_1 \alpha_i p_k^T p_k - c_1 \alpha_i (p_k - \nabla f_k)^T p_k
\end{equation}
and then approximate $(p_k - \nabla f_k)$ with
\begin{equation}
-2\cdot \nabla (\tr (R_i^{-1} R)+ \tr(R^{-1} \tilde R)).
\end{equation}

\section{Inflation of correlation estimates}\label{sec:inflation}

We provide here an example of our correlation estimation procedure which produces inflation in some unpenalized elements of the correlation matrix.
We solved the minimization problem in (\ref{eqn:minProblem}) with three different methods, finding identical answers each time, up to small numerical tolerances. Those methods are:
\begin{enumerate}
\item Estimate $R$ using our code, which appeals to the generalized gradient descent algorithm.
\item Estimate $R$ using a black-box numerical optimization algorithm, which has access to the function we're minimizing, but not its derivative.
\item Estimate $R$ by finding an analytic expression for the gradient of the function we're minimizing, and solve for a point where the gradient is zero.
\end{enumerate}

One case in which inflation manifests if we take our evidence from the data to be given by
\begin{equation}
\tilde R = \left(
\begin{array}{ccc}
1 & 0.8 & 0.5 \\
0.8 & 1 & 0.1 \\
0.5 & 0.1 & 1
\end{array}
\right)
\end{equation}
and the penalty matrix by
\begin{equation}
P = \left(
\begin{array}{ccc}
0 & 0 & 1 \\
0 & 0 & 0 \\
1 & 0 & 0
\end{array}
\right).
\end{equation}
We denote the unknown true correlation matrix by
\begin{equation}
R=\left(
\begin{array}{ccc}
1 & \rho_1 & \rho_2 \\
\rho_1 & 1 & \rho_3 \\
\rho_2 & \rho_3 & 1
\end{array}
\right).
\end{equation}
We fix the regularization parameter at $\lambda=0.5$.
The problem is then to estimate the three parameters $\rho_1$, $\rho_2$, and $\rho_3$. 

With all three methods we find an estimate of
\begin{equation}
\hat \brho = 
\left(
\begin{array}{c}
\hat \rho_1 \\ \hat \rho_2 \\ \hat \rho_3
\end{array}
\right)
=
\left(
\begin{array}{c}
0.8211 \\ 0.1542 \\ -0.1813
\end{array}
\right).
\end{equation}

Note that the second element, which is penalized, experiences shrinkage towards zero, as expected. The first element is inflated, while the third is both inflated and changes sign.

% AOS,AOAS: If there are supplements please fill:
%\begin{supplement}[id=suppA]
%  \sname{Supplement A}
%  \stitle{Title}
%  \slink[doi]{10.1214/00-AOASXXXXSUPP}
%  \sdatatype{.pdf}" 
%  \sdescription{Some text}
%\end{supplement}

\end{document}